\documentclass[letter, 10pt]{IEEEtran}

\usepackage{multirow}
\usepackage{cite}
\usepackage{amsmath,amssymb,amsfonts,amsthm}
\usepackage{algorithm}
\usepackage{algorithmic}
\usepackage{graphicx}
\usepackage{textcomp}
\usepackage{xcolor}
\usepackage{mathtools}
\usepackage{graphicx}
\usepackage{booktabs}
\usepackage{makecell}
\usepackage[font={small}]{caption}
\usepackage{subcaption}
\usepackage{cite}
\usepackage{breqn}
\usepackage{mathrsfs}
\usepackage{accents}
\usepackage{acronym}
\usepackage{multirow}
\usepackage{soul}
\usepackage{bm}
\usepackage{url}
\usepackage{wrapfig}
\usepackage{todonotes}
\usepackage{verbatim}
\usepackage{tikz}
\usepackage{pdfpages}
\usepackage{multirow}

\renewcommand{\thetable}{\arabic{table}}

\DeclareMathOperator*{\argmin}{arg\,min}

\graphicspath{{./figures/}}
\acrodef{prop}[\textit{MIMORPH}]{MIMO Radio Platform for Heterogeneous wireless systems}

\acrodef{abft}[A-BFT]{Association Beamforming Training}
\acrodef{ack}[ACK]{Acknowledge}
\acrodef{adc}[ADC]{Analog-to-Digital Converter}
\acrodef{aoa}[AoA]{Angle of Arrival}
\acrodef{aod}[AoD]{Angle of Departure}
\acrodef{ap}[AP]{Access Point}
\acrodef{amc}[AMC]{Advanced Mezzanine Card}
\acrodef{awv}[AWV]{Antenna Wave Vector}
\acrodef{axi}[AXI]{Advanced eXtensible Interface}
\acrodef{ber}[BER]{Bit Error Rate}
\acrodef{bft}[BFT]{Beamforming Training}
\acrodef{bp}[BP]{Beam Pattern}
\acrodef{brp}[BRP]{Beam Refinement Phase}
\acrodef{cfr}[CFR]{Channel Frequency Response}
\acrodef{cs}[CS]{Compressed Sensing}
\acrodef{cdf}[CDF]{Cumulative Distribution Function}
\acrodef{cef}[CEF]{Channel Estimation Field}
\acrodef{cfo}[CFO]{Carrier Frequency Offset}
\acrodef{cir}[CIR]{Channel Impulse Response}
\acrodef{csi}[CSI]{Channel State Information}
\acrodef{csirs}[CSI-RS]{CSI-Reference Signal}
\acrodef{cs}[CS]{Compressed Sensing}
\acrodef{cv}[CV]{Constant Velocity}
\acrodef{cnn}[CNN]{Convolutional Neural Network}
\acrodef{cots}[COTS]{Commercial-Off-The-Shelf}
\acrodef{dft}[DFT]{Discrete Fourier Transform}
\acrodef{dl}[DL]{Deep Learning}
\acrodef{dma}[DMA]{Direct Memory Access}
\acrodef{dmg}[DMG]{Directional Multi Gigabit}
\acrodef{dti}[DTI]{Data Transfer Interval}
\acrodef{dql}[DQL]{Double Q Learning}
\acrodef{edmg}[EDMG]{Enhanced Directional Multi Gigabit}
\acrodef{ekf}[EKF]{Extended Kalman Filter}
\acrodef{elu}[ELU]{Exponential-Linear Unit}
\acrodef{fmcw}[FMCW]{Frequency-Modulated Continuous-Wave}
\acrodef{fov}[FOV]{Field-of-View}
\acrodef{ft}[FT]{Fourier Transform}
\acrodef{gpio}[GPIO]{General Purpose Input/Output}
\acrodef{gsps}[GSPS]{Giga-Samples per Second}
\acrodef{har}[HAR]{Human Activity Recognition}
\acrodef{ht}[HT]{High Throughput}
\acrodef{if}[IF]{Intermediate Frequency}
\acrodef{ifs}[IFS]{Inter-Frame Spacing}
\acrodef{iht}[IHT]{Iterative Hard Thresholding}
\acrodef{ista}[ISTA]{Iterative Shrinkage-Thresholding Algorithm}
\acrodef{isac}[ISAC]{Integrated Sensing And Communication}
\acrodef{jcs}[JCS]{Joint Communication and Sensing}
\acrodef{jpdaf}[JPDAF]{Joint Probabilistic Data Association Filter}
\acrodef{los}[LOS]{Line-of-Sight}
\acrodef{lbm}[LBM]{Loop-Back Memory}
\acrodef{mae}[MAE]{Mean Absolute Error}
\acrodef{mcs}[MCS]{Modulation and Coding Scheme}
\acrodef{md}[mD]{micro-Doppler}
\acrodef{mimo}[MIMO]{Multiple Input Multiple Output}
\acrodef{bench1}[MC-SFS]{Mutual Coherence Sequential Forward Selection}
\acrodef{bench2}[RMC-RS]{Restricted Mutual Coherence Random Search}
\acrodef{mmwave}[mmWave]{Millimeter-Wave}
\acrodef{msps}[MSPS]{Mega-Samples per Second}
\acrodef{mu}[MU]{Multiple User}
\acrodef{MUSIC}[MUSIC]{MUlti SIgnal Classification}
\acrodef{mc}[MC]{Mutual Coherence}
\acrodef{mdp}[MDP]{Markov Decision Process}
\acrodef{mse}[MSE]{Mean Squared Error}
\acrodef{nac}[NAC]{Normalized Auto Correlation}
\acrodef{nco}[NCO]{Numerical Controlled Oscillator}
\acrodef{nlos}[NLOS]{Non-Line-of-Sight}
\acrodef{nn}[NN]{Neural Network}
\acrodef{ofdm}[OFDM]{Orthogonal Frequency Division Multiplexing}
\acrodef{omp}[OMP]{Orthogonal Matching Pursuit}
\acrodef{per}[PER]{Packet Error Rate}
\acrodef{phy}[PHY]{Physical Layer}
\acrodef{pl}[PL]{Programmable Logic}
\acrodef{pov}[POV]{Point-of-View}
\acrodef{ps}[PS]{Processing System}
\acrodef{ppo}[PPO]{Proximal Policy Optimization}
\acrodef{rf}[RF]{Radio Frequency}
\acrodef{rfsoc}[RFSoC]{Radio Frequency System on a Chip}
\acrodef{relu}[ReLU]{Rectified Linear Unit}
\acrodef{rl}[RL]{Reinforcement Learning}
\acrodef{rcs}[RCS]{Radar Cross-Section}
\acrodef{rss}[RSS]{Received Signal Strength}
\acrodef{rom}[ROM]{Read Only Memories}
\acrodef{sc}[SC]{Single Carrier}
\acrodef{sdr}[SDR]{Software Defined Radio}
\acrodef{siso}[SISO]{Single Input Single Output}
\acrodef{sls}[SLS]{Sector Level Sweep}
\acrodef{snr}[SNR]{Signal-to-Noise Ratio}
\acrodef{soc}[SoC]{System on a Chip}
\acrodef{spb}[SPB]{Signal Processing Blocks}
\acrodef{srrc}[SRRC]{Square-Root-Raised-Cosine}
\acrodef{ssb}[SSB]{Synchronization Signal Block}
\acrodef{ssr}[SSR]{Super Sample Rate}
\acrodef{sta}[STA]{Station}
\acrodef{stf}[STF]{Short Training Field}
\acrodef{stft}[STFT]{Short Time Fourier Transform}
\acrodef{su}[SU]{Single User}
\acrodef{tf}[TF]{Time-Frequency}
\acrodef{toa}[ToA]{Time of Arrival}
\acrodef{ula}[ULA]{Uniform Linear Array}
\acrodef{usrp}[USRP]{Universal Software Radio Peripheral}
\acrodef{vht}[VHT]{Very High Throughput}
\acrodef{wlan}[WLAN]{Wireless Local Area Network}
\acrodef{mab}[MAB]{Multi Armed Bandit}

\newcommand{\eq}[1]{Eq.~\eqref{#1}}
\newcommand{\fig}[1]{Fig.~\ref{#1}}
\newcommand{\tab}[1]{Tab.~\ref{#1}}

\newcommand{\rev}[1]{{\color{blue}#1}} 
\renewcommand{\rev}{}

\newcommand{\mytexttilde}{{\raise.17ex\hbox{$\scriptstyle\mathtt{\sim}$}}}

\IEEEaftertitletext{\vspace{-1.8\baselineskip}}

\begin{document}

\title{\LARGE Using Deep Reinforcement Learning to Enhance Channel Sampling Patterns in Integrated Sensing and Communication}

\author{Federico Mason, \IEEEmembership{Member,~IEEE}, Jacopo Pegoraro, \IEEEmembership{Member,~IEEE}
\thanks{The authors are with the Department of Information Engineering at the University of Padova, Italy (email: \texttt{<name>.<surname>@unipd.it}). This work was funded by the European Union under the Italian National Recovery and Resilience Plan (NRRP) Mission 4, Component 2, Investment 1.3, CUP C93C22005250001, partnership on “Telecommunications of the Future” (PE00000001 - program “RESTART”).}
}

\maketitle

\begin{abstract}
In \ac{isac} systems, estimating the \ac{md} spectrogram of a target requires combining channel estimates retrieved from communication with ad-hoc sensing packets, which cope with the sparsity of the communication traffic.
Hence, the \ac{md} quality depends on the transmission strategy of the sensing packets, which is still a challenging problem with no known solutions. 
In this letter, we design a deep \ac{rl} framework that fragments such a problem into a sequence of simpler decisions and takes advantage of the \ac{md} temporal evolution for maximizing the reconstruction performance. 
Our method is the first that learns sampling patterns to directly optimize the \ac{md} quality, enabling the adaptation of \ac{isac} systems to variable communication traffic. 
We validate the proposed approach on a dataset of real channel measurements, reaching up to 40\% higher \ac{md} reconstruction accuracy and several times lower computational complexity than state-of-the-art methods. 
\end{abstract}

\begin{IEEEkeywords}
Integrated sensing and communication, compressed sensing, micro-Doppler, reinforcement learning
\end{IEEEkeywords}

\IEEEpeerreviewmaketitle

\acresetall

\section{Introduction}

In \ac{isac}, sensing must coexist with primary communication functionalities while introducing minimal overhead and channel occupation. 
Jointly allocating resources for both communication and sensing tasks requires modifying the communication system~\cite{dong2022sensing}, which entails high implementation costs on existing wireless network devices.
A more affordable approach, termed \textit{communication-centric}, considers communication resources as pre-determined (i.e., non-modifiable), and \textit{reuses} them for sensing purposes, utilizing channel estimates obtained from communication packets~\cite{pegoraro2022sparcs}.
\rev{However, communication traffic patterns are typically irregular and sparse in time, due to the channel access and packet transmission protocols. This makes communication signals unsuited for traditional radar sensing applications, which require estimating the channel for extended periods at regular intervals~\cite{vandersmissen2018indoor}.}

A widely used example of such applications is the estimation of the \ac{md} spectrogram of a target. \rev{The latter is a frequency modulation of the received signal due to the movement of different parts of an extended target and enables object recognition and human sensing~\cite{zhang2021enabling}.}
To recover the \ac{md} from communication packets, \ac{cs} techniques are usually employed, exploiting the sparsity of the channel in the Doppler domain~\cite{eldar2012compressed}.
However, when communication traffic is highly irregular, even \ac{cs} has been shown to provide degraded performance~\cite{pegoraro2022sparcs}.
In such cases, the available channel estimates (or \textit{samples}) obtained from communication packets can be integrated with \textit{sensing-oriented} transmissions involving only packet headers, as first suggested in~\cite{pegoraro2022sparcs}.
Transmitting sensing packets at suitable times is key to maximizing the \ac{md} quality.
However, finding optimal sampling times is very challenging due to the complex relation between the sampling pattern and the \ac{cs} performance.

In radar systems there is full control over the signal transmission times and channel sampling patterns are usually obtained by minimizing the \ac{mc} of the \ac{cs} model matrix~\cite{xu2014compressed}.
However, \ac{mc} minimization is a combinatorial problem, and an optimal solution can only be found for specific configurations of the system parameters~\cite{song2024nonuniform}.
Recently, greedy algorithms have been used to find approximate solutions~\cite{song2024nonuniform} but present high computational complexity that prevents their implementation in real-time.
Moreover, none of the state-of-the-art methods works when part of the sampling instants is fixed, a specific issue inherent to \ac{isac}.  
Hence, two main research challenges arise:
(i)~\ac{mc} minimization is computationally complex and does not imply high \ac{md} reconstruction quality, 
(ii)~in \ac{isac}, only \textit{part} of the channel sampling instants can be optimized making existing solutions for radars not directly applicable.  

In this letter, we tackle the above challenges by exploiting a deep \ac{rl} approach.
Specifically, we design a framework where the problem of allocating sensing packets is decomposed into a sequence of decisions made by a \ac{rl} agent.
The agent state includes the planned sampling pattern and the contextual information given by the \ac{md} reconstructed in the previous processing window.
At each decision step, the agent plans the transmission of new sensing packets and receives a reward that is directly dependent on the quality of the \ac{md} reconstruction.
This avoids the expensive computation of the \ac{mc}, significantly reducing the complexity of allocating sensing packets with respect to known methods.

We test the proposed framework on the DISC dataset~\cite{pegoraro2022disc}, which contains \ac{cir} measurements involving human movements.
We use the Proximal Policy Optimization (PPO) algorithm to train the agent and compare its performance against two \ac{mc} minimization algorithms from the recent literature.
Experimental results show that our solution obtains significantly better channel sampling patterns than \ac{mc}-based methods at a fraction of the computational time, demonstrating how data-driven approaches are highly beneficial to optimize \ac{isac} systems.

\section{Background and system model}\label{sec:model}

In this section, we present the \ac{isac} system where our framework is implemented, and provide background on \ac{cs}-based \ac{md} reconstruction.
The system is based on a \ac{cir} model adapted from~\cite{pegoraro2022sparcs}, which is compliant with the IEEE~802.11ay \ac{isac} implementation used in the experiments~\cite{pegoraro2022disc}.

\subsection{Channel impulse response model}\label{sec:cir-model}
The \ac{isac} system operates with processing windows of $W$ discrete timesteps, indexed by $w \in \mathcal{W} = \{ 0, \dots, W-1\}$.
The timestep duration is denoted by $T_c$ and is assumed short enough that the target Doppler spectrum is constant within a processing window.
The \ac{cir} $h(\tau, wT_c)$ associated with the window depends on the propagation delay $\tau$ and time $wT_c$,
and is expressed as a sum of $L$ Dirac delta components which correspond to the \textit{resolvable} signal propagation paths. 
Denoting by $\tau_l$ the propagation delay of the \mbox{$l$-th} reflector, and by $\tilde{h}_{l}[w]$ the \mbox{$l$-th} \ac{cir} component, we have 
\begin{equation}\label{eq:cir-continuous}
    h(\tau, wT_c) = \sum_{l=1}^{L} \tilde{h}_{l}[w] \delta(\tau - \tau_l),
\end{equation}
where we used parentheses $(\cdot)$ and square brackets $[\cdot]$ to denote continuous and discrete-time domains, respectively. 
Assuming that the multipath parameters are constant within a processing window, we have
\begin{equation}\label{eq:cir-simple}
    \tilde{h}_l[w] =\sum_{q=1}^{Q_l} \alpha_{l, q}  e^{j2\pi f_{l, q} w T_c},
\end{equation}
where $\alpha_{l, q}$ and $f_{l, q}$ are the complex amplitude and Doppler frequency of the $q$-th scatterer contributing to the $l$-th resolvable path, respectively.
Therefore, each \ac{cir} component is expressed as a superposition of $Q_l$ complex exponential terms with frequencies $f_{l, q}$. 
We observe that the \ac{cir}, for a specific delay $\tau_l$, is sparse in the Doppler frequency domain, since $Q_l \ll W$ for typical processing window lengths.

\rev{We assume that the \ac{isac} system can detect and track the targets of interest, extracting the corresponding paths from \eq{eq:cir-continuous} by estimating $\tau_l$ for $l=1, \dots, L$.
This can be done, e.g., by peak detection followed by Kalman filtering~\cite{pegoraro2023rapid}. Hence, our algorithm treats each target independently, processing the \ac{cir} estimates for the corresponding path. For this reason, under the above assumption, the distinction among different paths does not pertain to the extraction of \ac{md}, so we simplify the notation by dropping the path index $l$, thus writing $\tilde{h}[w]$ to denote a \ac{cir} sample.}

Our focus is reconstructing the \ac{md} spectrogram of the target from the Doppler domain channel $\mathbf{H}$, which corresponds to the \ac{dft} of the \ac{cir} over the processing window.
\rev{Particularly, if all the \ac{cir} samples in a window were available, $\mathbf{H}$ could be obtained as $\mathbf{H} = \mathbf{F}^{-1}\mathbf{g}$~\cite{vandersmissen2018indoor}, where $\mathbf{F}$ is the inverse Fourier matrix of dimension $W$, whose elements are \mbox{$F_{nw} = e^{j 2\pi nw / W}/\sqrt{W}, \,\, n,w \in \mathcal{W}$}, while $\mathbf{g}=[\tilde{h}[0], \dots, \tilde{h}[W-1]]^{\mathsf{T}}$ is the vector containing the \ac{cir} samples. However, \ac{isac} systems do not typically transmit packets at a regular rate due to the communication protocol~\cite{pegoraro2022sparcs}. In the next section, we modify the problem formulation, generalizing it to the case of irregular \ac{cir} estimates, which can be modeled as missing samples from vector $\mathbf{g}$.}




\subsection{micro-Doppler sparse recovery problem}\label{sec:md-sparse-rec}
In \ac{isac} systems, only a subset of \ac{cir} samples is available, since \ac{cir} estimation is performed (i)~for communication packets, which enable channel equalization, and (ii)~for sensing-only packets, which only contain channel estimation fields. 
While the scheduling of sensing packets can be personalized, communication packets are sent at irregular times.
Hence, reconstructing the \ac{md} spectrogram from communication packets only would result in severe artifacts that strongly degrade the system performance. 

Given the sparsity of the \ac{cir} in the Doppler domain, \ac{md} reconstruction can be cast as a sparse recovery problem.
\rev{To model irregular \ac{cir} estimation times, we assume that \ac{cir} estimates are only obtained at a subset of $M$ instants out $W$ time-steps in the window, whose indices are collected in set $\mathcal{M} = \{i_m\}_{m=1}^{M}$.}
It holds that $\mathcal{M} = $ $\mathcal{M}_c \cup \mathcal{M}_s$, where $\mathcal{M}_c$ contains the \ac{cir} samples associated with communication packets, as specified in point (i) above, and $\mathcal{M}_s$ includes the sensing samples (point (ii)).

We now denote by $\mathbf{M}=\left[ \mathbf{e}^{\mathsf{T}}_{i_m}\right]$, $\forall \, i_m \in \mathcal{M}$, the matrix that selects the rows of $\mathbf{F}$ whose indices are in $\mathcal{M}$.
Specifically, $\mathbf{e}_{i}$ is the $i$-th element of the canonical basis, i.e., the vector of all zeros but the $i$-th component, which equals $1$.
Calling $\mathbf{h}=\left[\tilde{h}[i_1], \dots, \tilde{h}[i_M]\right]^{\mathsf{T}}$ the vector containing the available \ac{cir} samples, we can write
$\mathbf{h}  = \mathbf{\Psi} \mathbf{H} + \mathbf{n}$,
where \mbox{$\mathbf{\Psi}=\mathbf{M}\mathbf{F}$} is the sensing matrix and $\mathbf{n}$ is a noise vector. 

A reconstruction of $\mathbf{H}$ can be obtained by solving the following \ac{cs} problem
\begin{equation}\label{eq:cs-iht}
    \hat{\mathbf{H}}= \argmin_{\mathbf{H}} ||\mathbf{h} - \mathbf{\Psi}\mathbf{H}||_2^2 \, \mbox{ s.t. } ||\mathbf{H}||_0  \leq \Omega,
\end{equation}
where $\Omega$ is the so-called \textit{sparsity level} and determines the maximum number of non-zero components of $\mathbf{H}$.
The above problem can be solved
using iterative algorithms such as \ac{iht} or \ac{omp}~\cite{eldar2012compressed}.
Once $\hat{\mathbf{H}}$ has been obtained, the \ac{md} spectrum is computed as $\hat{\mathbf{D}} = |\hat{\mathbf{H}}|^2$.


\subsection{Mutual coherence minimization}

The most popular approach to designing sampling patterns that optimize \ac{cs} reconstruction is to minimize the \ac{mc} of the sensing matrix.
The \ac{mc} is defined as 
\begin{equation}\label{eq:coherence}
    \mu(\mathbf{\Psi}) = \max_{ \substack{ i,l = 0, \dots, K-1 \\  i\neq l}} | \mathbf{\Psi}_{i}^{\mathsf{H}}\mathbf{\Psi}_{l}|,
\end{equation}
where $\mathbf{\Psi}_{i}$ and $\mathbf{\Psi}_{l}$ are the $i$-th and $l$-th columns of $\mathbf{\Psi}$, and $\mathbf{X}^{\mathsf{H}}$ represents the Hermitian of matrix $\mathbf{X}$.
The \ac{mc} measures the maximum correlation among the columns of $\mathbf{\Psi}$. Lower \ac{mc} corresponds to better reconstruction performance using \ac{cs} algorithms~\cite{xu2014compressed}.
In general, \ac{mc} decreases as more sensing samples are used for reconstructing the \ac{cir} spectrum, i.e. when $M$ increases.
However, the minimization of the \ac{mc} is a combinatorial problem \cite{xu2014compressed} for which an optimal solution can only be found for specific pairs of $W$ ad $M$~\cite{song2024nonuniform}.
In our scenario, $W$ depends on the specific \ac{isac} system, while $M$ varies in time. Therefore, finding an optimal sampling pattern via \ac{md} minimization is infeasible, and approximate solutions are needed.
Moreover, the \ac{isac} setting introduces the following two technical challenges. 

\noindent\textit{1) Discrepancy between \ac{mc} and sensing performance.}
The quality of the \ac{cs} reconstruction can be quantified by the \ac{mse} between the reconstructed \ac{md} and that obtained from a complete \ac{cir} sampling window.
However, an explicit relation between the \ac{mc} and the \ac{mse} has not been proven and, as a result, a costly minimization of the \ac{mc} could yield limited performance improvement.
To achieve substantially better sensing performance, it would be desirable to select sampling patterns to \textit{directly} enhance the \ac{mse}.

\noindent\textit{2) Partial control over sampling patterns.}
The \ac{cir} samples obtained from communication packets ($\mathcal{M}_c$) are not under control and their collection times are fixed.
This means that $\mathcal{M}$ can not be selected arbitrarily and only the locations of samples in $\mathcal{M}_s$ can be scheduled to optimize \ac{mc}.
We stress that this is made even more challenging by the fact that 
the optimization of the sensing samples has to be \textit{dynamically} carried out during system operation.


\section{Methodology}\label{sec:method}
In the following, we design a \ac{rl} framework to select the sampling pattern in the target \ac{isac} scenario.
Our framework assumes that the communication packets' transmission instants become available at the start of each processing window.
Hence, a \ac{rl} agent has assigned the task of building the set $\mathcal{M}_s$ with the final goal of minimizing the error between the true \ac{md} spectrum and the one reconstructed from the samples in $\mathcal{M}_s$ by using the \ac{iht} algorithm~\cite{eldar2012compressed}.

\subsection{Reinforcement learning model}
\label{sub:reinf}

We model the \ac{isac} system as a \ac{mdp} encoded by the tuple $\left( \mathcal{S}, \mathcal{A}, R, P, \gamma \right)$~\cite{sutton2018reinforcement}.
In particular, $\mathcal{S}$ is the state space, $\mathcal{A}$ is the action space, $R: \mathcal{S} \times \mathcal{A} \times \mathcal{S} \rightarrow \mathbb{R}$ is the reward function, $P: \mathcal{S} \times \mathcal{A} \times \mathcal{S} \rightarrow [0,1]$ is the transition probability function, and $\gamma \in [0,1)$ is the \emph{discount factor} that trades off between foresighted and myopic policies. 

We consider that the processing of each window constitutes a distinct learning episode.
At the start of the episode, the set $\mathcal{M}_s$ is empty since no transmissions of sensing packets have been planned. 
Hence, each agent's action inserts a new sample within $\mathcal{M}_s$, and the agents continue to choose actions until the size of $\mathcal{M}_s$ is equal to $M -  |\mathcal{M}_c|$.
We observe that the episode duration is variable and depends on the cardinality of $\mathcal{M}_c$, i.e., the number of communication packets that will be transmitted within the processing window.

We denote by $k=0, 1, ..., K-1$ the time steps within the same episode, with $K = M -  |\mathcal{M}_c|$.
The state $s[k] \in \mathcal{S}$ observed at time $k$ is modeled as a tuple $(\hat{\mathbf{H}}_{\text{prev}}, \mathcal{M}[k])$, where $\hat{\mathbf{H}}_{\text{prev}}$ is the Doppler-domain channel obtained at the end of the previous episode, while $\mathcal{M}[k]$ is the set $\mathcal{M}$ after $k+1$ agent decisions.
This makes the state representation include two different components: $\hat{\mathbf{H}}_{\text{prev}}$, which does not vary in time, and $\mathcal{M}[k]$, which changes as the agent takes new decisions. 

We remark that the choice to model the sample selection as a \textit{sequential} decision-making problem is non-trivial and leads to significant advantages. A more intuitive, but na{\"i}ve, approach is to use contextual \ac{mab}~\cite{sutton2018reinforcement}, where $\hat{\mathbf{H}}_{\text{prev}}$ represents the \textit{context} observed by the agent and the action space is given by all the possible combinations of $\mathcal{M}$. However, this approach shares the same limitations of \ac{mc} minimization, namely it requires performing optimization over a combinatorial space. Indeed, the \ac{mab} action space contains all the possible combinations of sample instants selection over the available slots in the window, which has size $(W-|\mathcal{M}_c|)! / \left( (M-|\mathcal{M}_c|)! (W-M)! \right)$.



Unlike \ac{mab}, our framework upper bounds the cardinality of the action space by $W-|\mathcal{M}_c|$.
Indeed, the agent's action space at slot $k$ includes the indexes of all the idle samples so that $\mathcal{A}[k]= \mathcal{W} \setminus \mathcal{M} [k]$.
If the agent chooses action $a$, the latter is added to the set $\mathcal{M}$ and the environment evolves toward the state $s[k+1]=(\hat{\mathbf{H}}_{\text{prev}}, \mathcal{M}[k+1])$, where $\mathcal{M}[k+1] = \mathcal{M}[k] \cup \{ a \}$.
This makes the \ac{mdp} transitions deterministic, i.e., $s[k+1]$ is fully predictable given $s[k]$ and $a[k]$. 

Contrary to the transition probability function $P(\cdot)$, the reward function $R(\cdot)$ presents a stochastic behavior. 
Indeed, we make the reward depend on the true \ac{cir} spectrum, which is not part of the system state and whose reconstruction is the goal of the learning agent. 
Let us denote by $\hat{\mathbf{H}}[k]$ the estimation of the \ac{cir} spectrum that is obtained with \ac{iht} using $\mathcal{M}[k]$, i.e., all the samples selected up to time $k$. 
At step $k$, if the agent chooses action $a[k]$, the reward $r[k] = R(s[k], a[k], s[k+1])$ is equal to
\begin{equation}
    r[k]=\left({\rm MSE}(\hat{\mathbf{H}}[k+1], \mathbf{H}) - {\rm MSE}(\hat{\mathbf{H}}[k], \mathbf{H}) \right)\Big/ \lVert\mathbf{H}\rVert_2^2,
     \label{eq:reward}
\end{equation}
where ${\rm MSE}\left(\mathbf{x}, \mathbf{y}\right) = \sum_{i=0}^{n-1} (x_i - y_i) ^ 2 / n$ is the \ac{mse} between vectors $\mathbf{x}$ and $\mathbf{y}$, while $\lVert\mathbf{x}\rVert_2$ is the Euclidean norm of $\mathbf{x}$.

In our system, the reward increases as the agent selects more useful samples for spectrum estimation.
The agent decisions are determined by a policy $\pi: \mathcal{S} \times \mathcal{A} \rightarrow [0, 1]$, where $\pi(s, a)$ represents the probability of taking action $a$ in state $s$.
Hence, the agent's goal is to find the policy that maximizes the reward earned over each episode, given by the discounted return $G[k] = \sum_{\nu=k}^{K-1} \gamma^{\nu - k} r[\nu]$.
\rev{Finally, we stress that the reward, and hence the ground-truth Doppler domain channel $\mathbf{H}$, is only used to \textit{train} the agent while, during the inference, our system does not require the knowledge of $\mathbf{H}$.}

\subsection{Agent architecture and training}

To make the agent find the optimal strategy, we use \ac{ppo}, a \emph{policy gradient algoirthm} that allows us to concurrently estimate both the policy $\pi$, as defined in the previous section, and the state-value function $V_\pi: \mathcal{S} \rightarrow \mathbb{R}$.
The latter associates each state $s$ with the expectation $E_{\pi}[G[k]|s[k]=s]$ of the cumulative return obtained given that the agent follows policy $\pi$. 
More details about \ac{ppo} can be found in the original paper~\cite{schulman2017proximal}.

To avoid the problems associated with the \emph{curse of dimensionality}, we implement both the policy $\pi(\cdot)$ and the state-value function $V_\pi(\cdot)$ as \rev{\ac{cnn} models.
Both models take as input the state $s$ that the agent observes while returning as output a single value, in the case $V_\pi(\cdot)$, and a probability distribution over the action space, in the case of $\pi(\cdot)$. 
The hidden architecture of the two models is the same and includes a sequence of $4$ one-dimensional convolutions with a fixed kernel size (equal to 5),  and an increasing channel number, which reaches the value of $32$ in the last layer. 
Both models alternate the hidden layers with rectified linear units as nonlinear activation functions and implement a feed-forward layer at the output.}
While the $V_\pi(\cdot)$ model ends with a simple linear combination, we add a \emph{softmax} function at the end of the $\pi$ model to ensure that the output corresponds to a probability distribution.

During the training phase, we make the agent interact with the environment for $N_{\text{train}}$ windows.
In doing so, we assume that $\hat{\mathbf{H}}_{\text{prev}}$, i.e., the spectrum reconstructed after processing the previous window, is identical to the ground truth so that $\hat{\mathbf{H}}_{\text{prev}} = \mathbf{H}$. 
This allows the agent to interact with an ideal environment at training time, with perfect information about the channel in the previous processing window.
Conversely, during the test phase, which lasts  $N_{\text{test}}$ episodes, $\hat{\mathbf{H}}_{\text{prev}}$ is obtained using the samples chosen by the agent during the previous episode, as occurs in a real \ac{isac} scenario.

\section{Settings and results}
\label{sec:results}

In this section, we validate our approach on real \ac{cir} traces and against existing methods from the literature.
\rev{To this goal, we exploit the public DISC dataset~\cite{pegoraro2022disc}, which contains $416$ IEEE~802.11ay \ac{cir} sequences, obtained using a monostatic \ac{isac} platform operating at $60$~GHz carrier frequency. The \ac{cir} is estimated at a fixed sampling rate of $T_c = 0.27$~ms, using pilot signals based on Golay sequences with $1.76$~GHz of bandwidth. No preprocessing is applied on the \ac{cir} estimates. For additional details on the experimental setting, we refer to~\cite{pegoraro2022disc}.}
The dataset contains signal reflections on $7$ subjects performing four different activities: \textit{walking}, \textit{running}, \textit{waving hands}, and \textit{sitting down/standing up}. 
\rev{We remark that the \ac{ppo} algorithm was trained directly on DISC, whose traces were preliminary split between training and testing sets.}

\subsection{Scenario settings}

In our experiments, we set the window size to $W=64$, and consider three possible cardinality for the set $\mathcal{M}$, i.e., $M \in \{8, 16, 32\}$.
We obtain incomplete \ac{cir} measurement patterns by randomly removing samples from the DISC \ac{cir} sequences.
To determine the samples associated with the transmissions of communication packets, we define a Markov chain with two states, named $\sigma_{i}$ and $\sigma_{t}$, that establishes if a certain slot stays idle or belongs to $\mathcal{M}_c$.
When a new window is processed, the first slot is assigned to any state according to the steady state probabilities, while the remaining slots are associated with $\sigma_{i}$ or $\sigma_{t}$ according to the chain evolution. 

We write $p_{i, t}$ and $p_{t, i}$ to denote the probability of going from state $\sigma_{i}$ to state $\sigma_{t}$, and vice versa.
These probabilities depend on two additional parameters, $b$ and $d$, and are defined as $p_{i, t} = d  p_{t, i} / (W - d)$, $p_{t, i} = 1 - 1 / b$, $p_{i, i} = 1 - p_{i, t}$ and $p_{t, t} = 1 - p_{t, i}$.
Hence, $b$ denotes the tendency of having a burst of consecutive communication slots, while $d$ is the density of communication slots within the same window.

We generate multiple Markov chain configurations by selecting $b \in \{M / 2, M, 2 M\}$ and $d \in \{M / 8, M / 4, M / 2\}$. 
For each configuration, we implement an ad hoc \ac{rl} agent trained for $3 \cdot 10^4$ episodes and setting $\gamma=0.99$. 
We exploit Adam as neural network optimizer~\cite{kingma2017adammethodstochasticoptimization}, considering $10^{-3}$ and $5\cdot10^{-5}$ as the maximum \textit{learning rate} for $\pi(\cdot)$ and $V_{\pi}(\cdot)$, respectively.
The \ac{ppo} algorithm was tuned by using  $0.1$  as \textit{clipping range} and $0.01$ as \textit{entropy coefficient}.
During the test, we analyze $10$ sequences of $250$ windows each, which sets the total number of testing episodes to $2500$. 

We compare our \ac{rl} framework against random sample selection and $2$ recent algorithms from the literature, namely \ac{bench1} and \ac{bench2}~\cite{song2024nonuniform}.
The random policy selects the transmission instants of sensing packets according to a uniform distribution over the available slots in the processing window.
Instead, \ac{bench1} follows a greedy approach to iteratively select the sample that minimizes the \ac{mc}.
Finally, \ac{bench2} exploits the last spectrum reconstruction, i.e., $\hat{\mathbf{H}}_{\text{prev}}$, to restrict the computation of the \ac{mc} to a subset of columns of the sensing matrix; then, it randomly generates a set of different sampling patterns and selects the one that minimizes the \ac{mc}.
We specify that \ac{bench1} and \ac{bench2} were designed for settings where the sampling pattern can be freely chosen, i.e., $\mathcal{M}_{\rm c} = \O$.
Therefore, we adapt \ac{bench1} and \ac{bench2} to our case by starting from the given set of samples $\mathcal{M}_c$ and applying them to obtain the total sampling pattern selecting the remaining $\mathcal{M}_s$ samples.

\subsection{micro-Doppler reconstruction quality}

\begin{figure}[t!]
	\begin{center}   
		\centering
		\includegraphics[width=0.49\columnwidth]{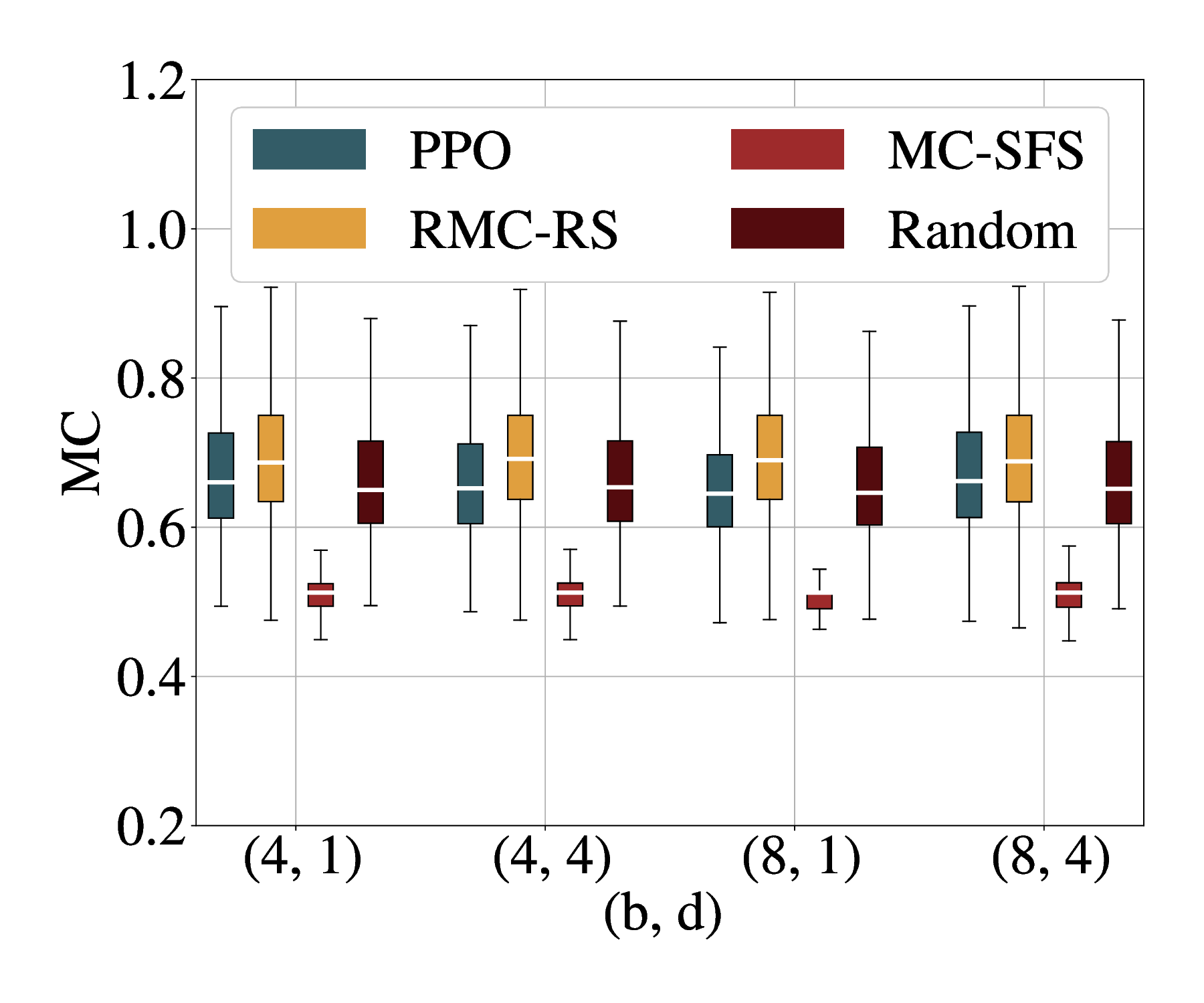}
		\includegraphics[width=0.49\columnwidth]{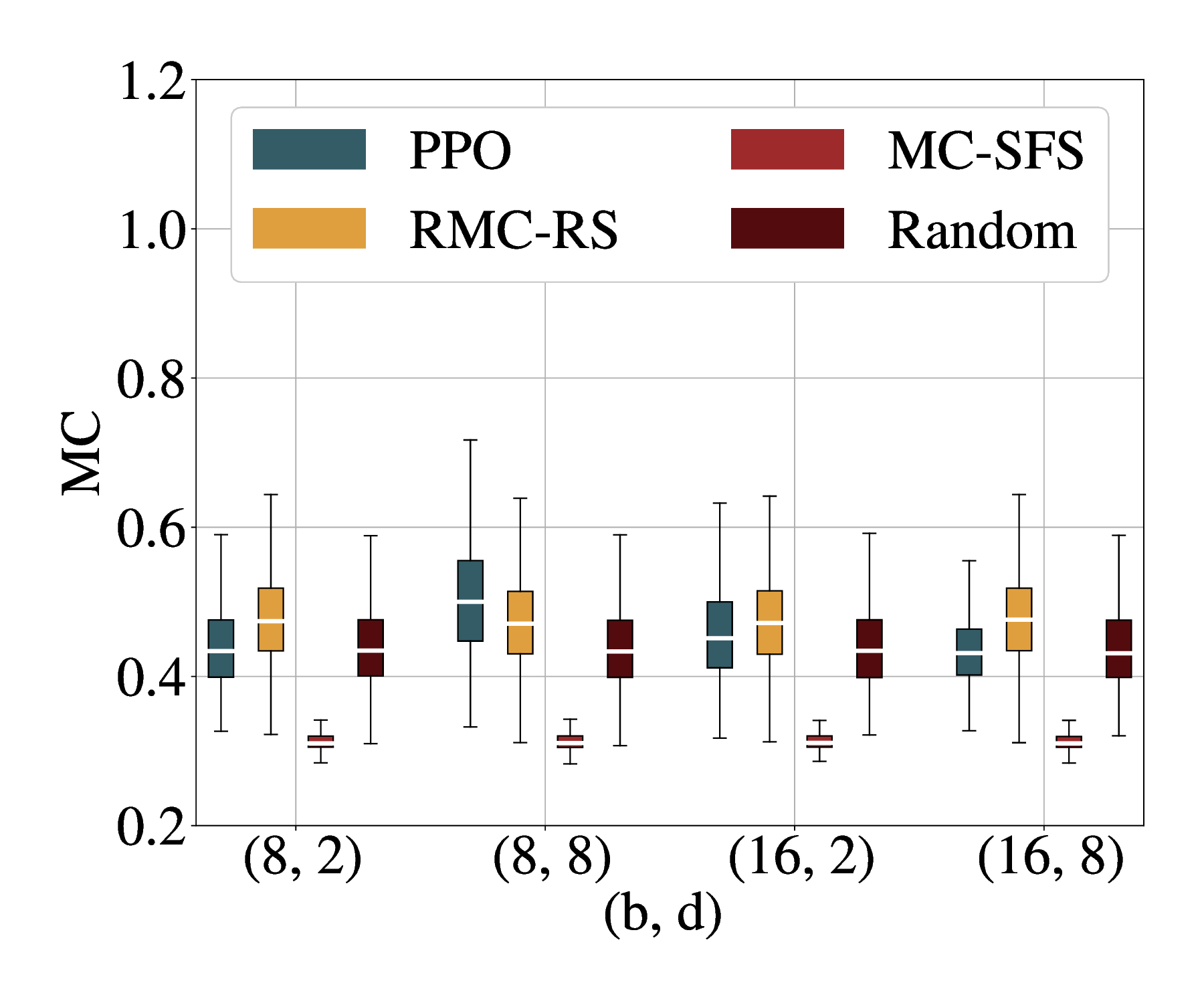}
		\caption{\acf{mc} for $M=8$ (left) and $M=16$ (right), considering $W=64$ as windows length.}
		\label{fig:coherence_64}
	\end{center}
  \vspace{-4mm}
\end{figure}

\begin{figure}[t!]
	\begin{center}   
		\centering
		\includegraphics[width=0.49\columnwidth]{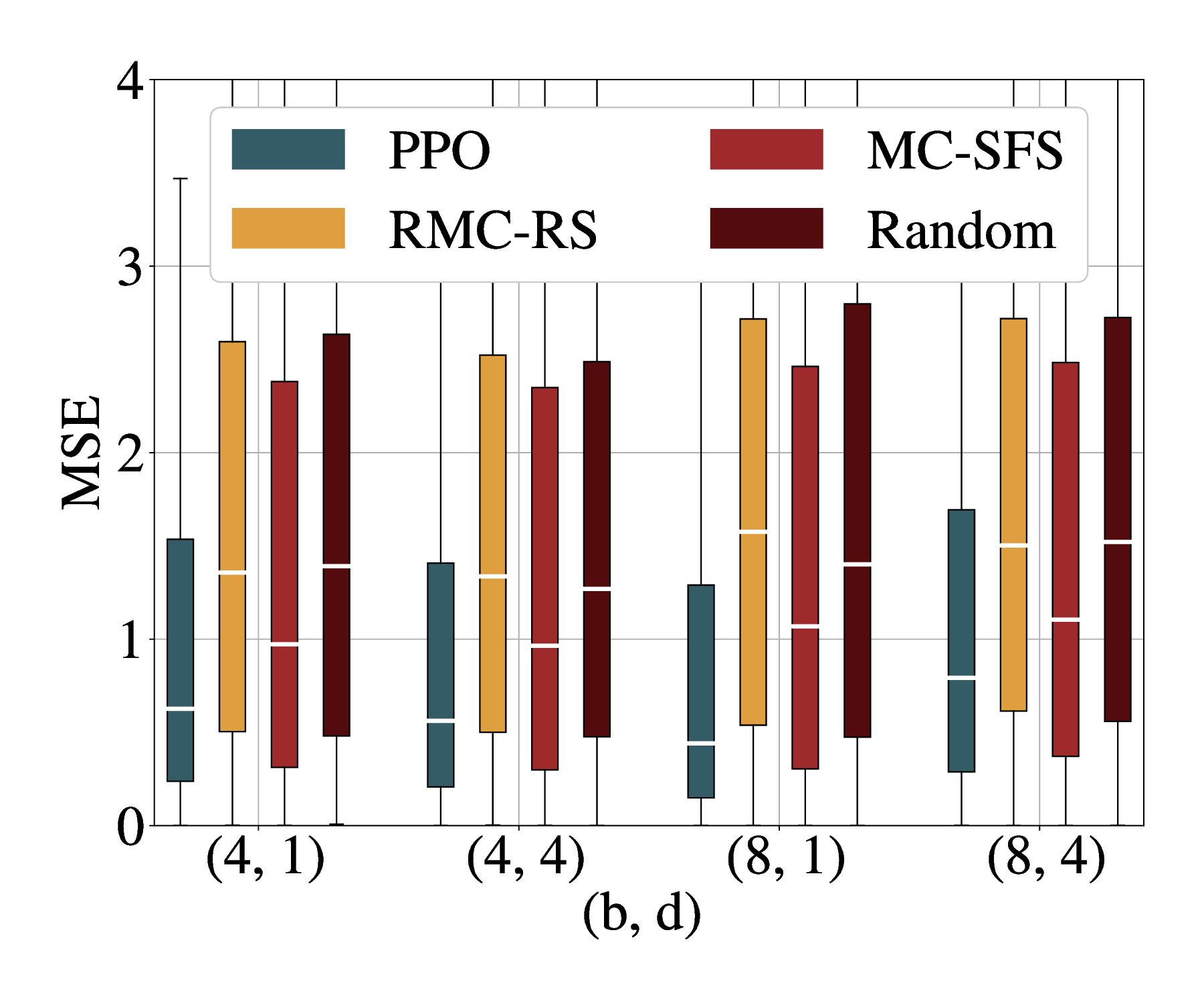}
		\includegraphics[width=0.49\columnwidth]{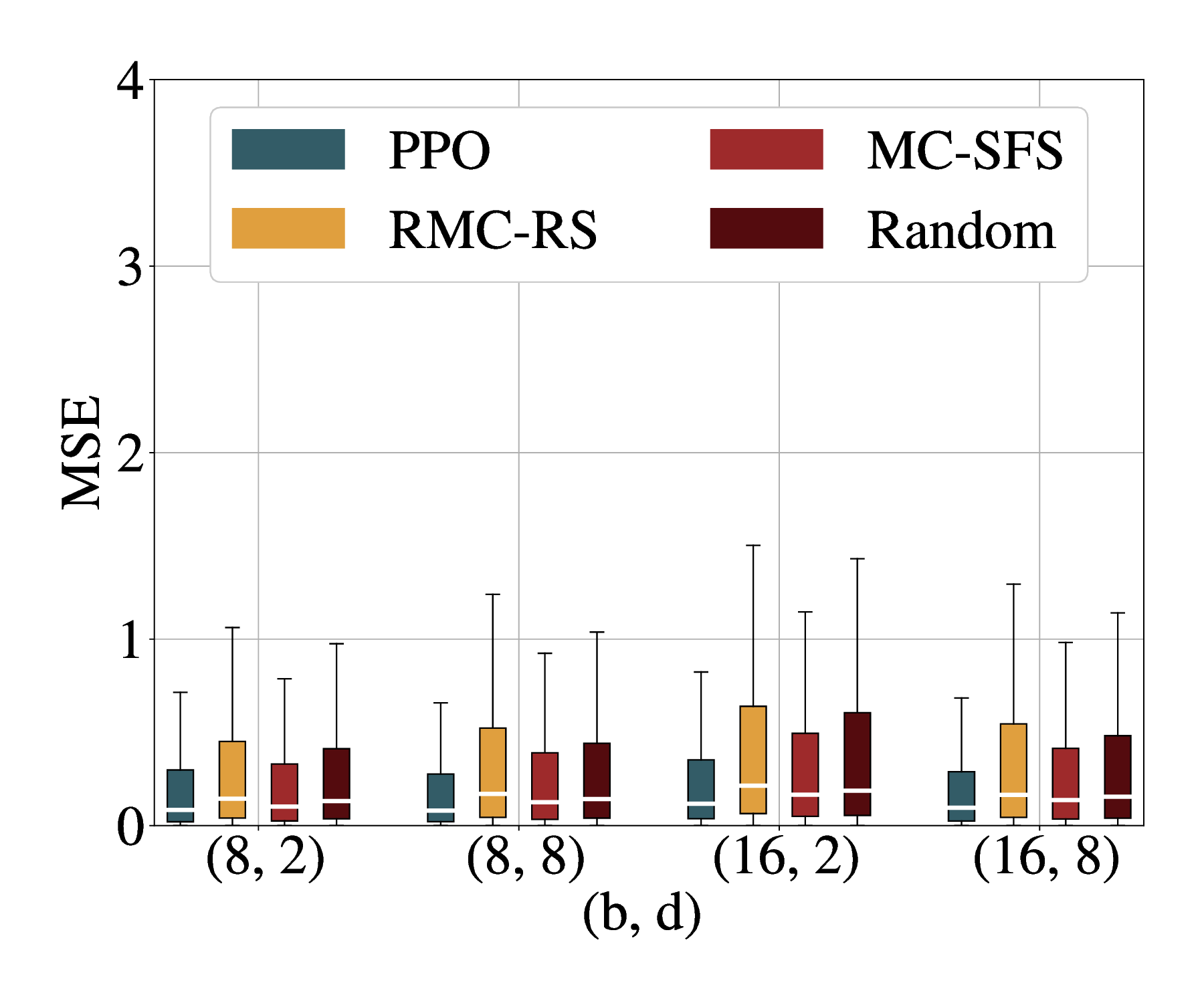}
		\caption{\acf{mse} for $M=8$ (left) and $M=16$ (right), considering $W=64$ as window length.}
		\label{fig:mse_64}
	\end{center}
  \vspace{-4mm}
\end{figure}
To assess the performance of the different techniques, we consider both the \ac{mc} and the \ac{mse} between the original and reconstructed spectrum. 
In Fig.~\ref{fig:coherence_64}, we represent the distribution of the \ac{mc} using a boxplot representation, according to different scenario configurations. 
We observe that \ac{bench1} leads to the best performance in terms of \ac{mc} in all the scenarios as it is explicitly designed to minimize it.
\ac{ppo} and random selection are agnostic to \ac{mc} and thus present similar \ac{mc} statistics.
\ac{bench2} leads to even higher \ac{mc} values, which may indicate that such a technique is ineffective in our setting, due to the fast variation of the \ac{md} spectrum across time. 

In terms of \ac{mse}, the proposed \ac{rl} strategy leads to the best results and, in some circumstances, improves the median \ac{mse} obtained with \ac{bench1} by more than $40\%$. 
Looking at the percentiles of the \ac{mse} distribution, the benefits of the \ac{ppo} algorithm are even more evident. 
For instance, considering $M=8$, $b=4$, and $d=1$, the $75$th percentile of the \ac{mse} distribution is about $1.5$ with \ac{ppo}, and $2.4$ with \ac{bench1}.
This confirms that optimizing \ac{mc} does not guarantee higher \ac{md} quality, encouraging the use of learning strategies trained on minimizing \ac{mse} on real data. 
\ac{bench2} leads to very similar results to those obtained with the random selection and it is outperformed by both \ac{bench1} and \ac{ppo} in all scenarios.

\begin{figure}[t!]
	\begin{center}   
		\centering
		\includegraphics[width=0.49\columnwidth]{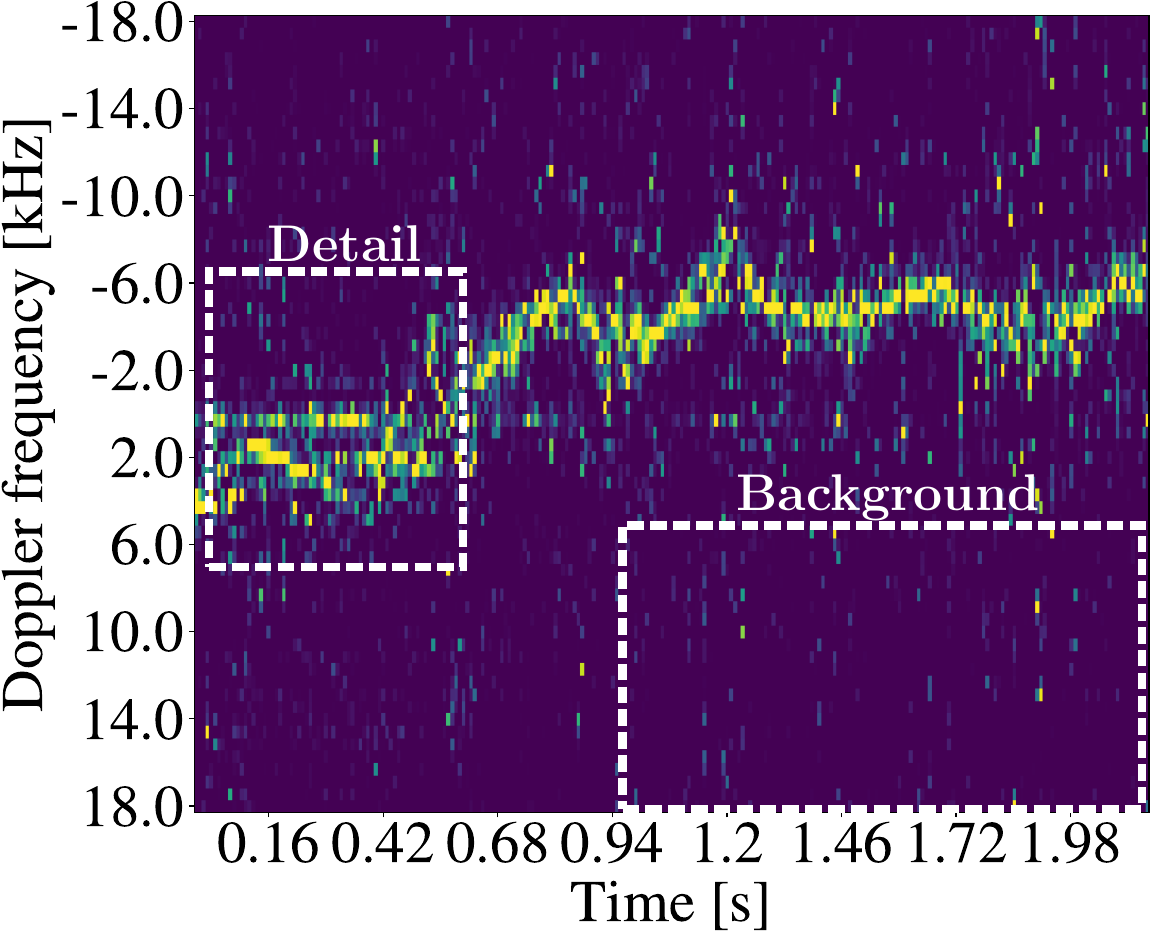}
		\includegraphics[width=0.49\columnwidth]{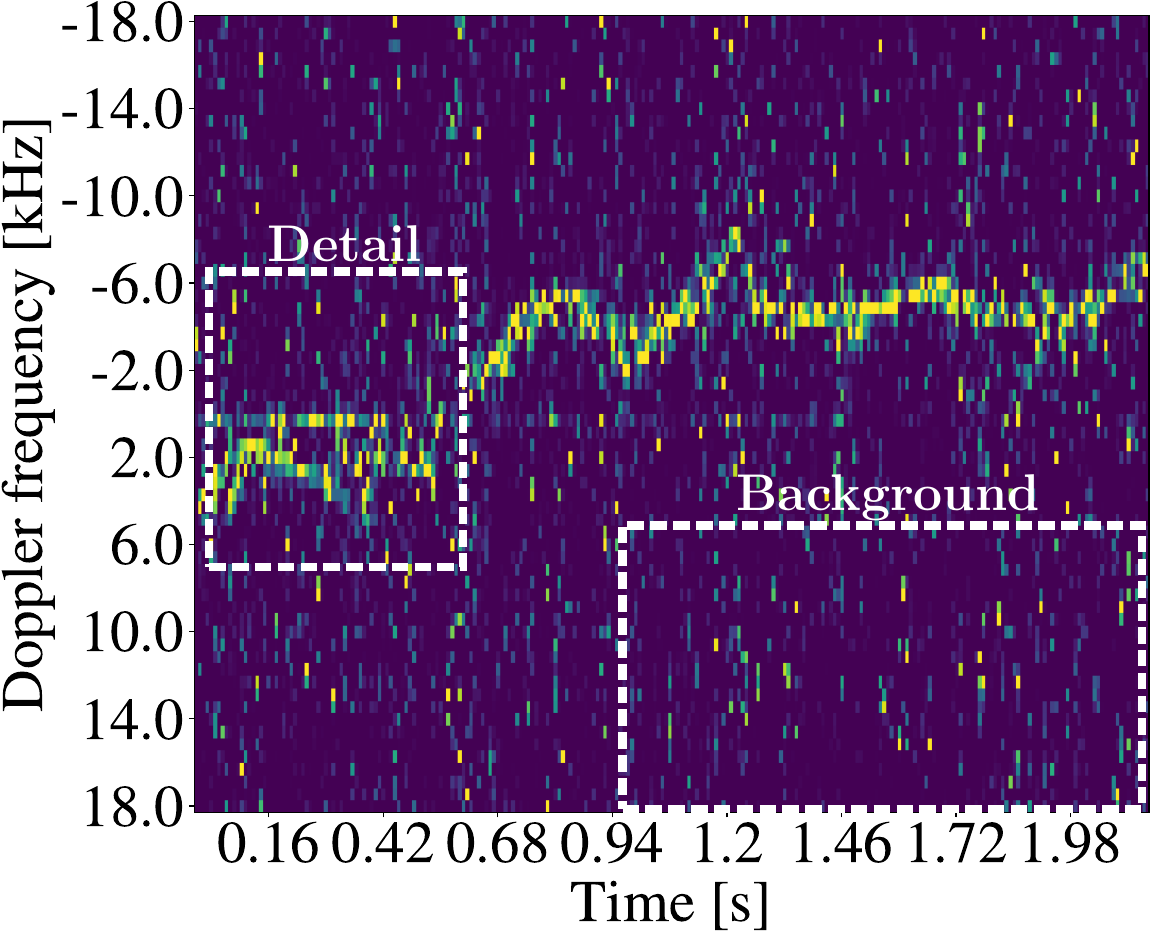}
		\caption{\rev{\textit{Walking} Doppler spectrum using PPO (left) and MC-SFS (right), with $(M,b,d)=(8,8,1)$. \ac{ppo} shows higher detail and less background noise than MC-SFS, as shown in the dashed boxes.}}
		\label{fig:spectrum_4}
	\end{center}
 \vspace{-3mm}
\end{figure}

We observe that varying the Markov chain statistics (i.e., the values of $b$ and $d$) does not substantially affect the results.
Conversely, increasing $M$ makes it possible to further reduce the \ac{mse} at the cost of transmitting more sensing packets.
\rev{Finally, to visualize the benefits of the proposed approach, in \fig{fig:spectrum_4}, we represent an example of Doppler spectrogram reconstructed by \ac{ppo} and \ac{bench1}, considering $M=8$, $b=8$, and $d=1$.
As highlighted in the dashed boxes, the \ac{md} obtained by \ac{ppo} shows higher detail, which is useful for distinguishing different components in the spectrogram, and is less affected by background noise.}

\subsection{Computational complexity}
\rev{In the following, we compare our approach to existing methods in terms of computational complexity.  The random selection strategy does not optimize the sample selection, so its complexity is the lowest among the considered methods, requiring $\mathcal{O}(M)$ operations.
As shown in~\cite{song2024nonuniform}, \ac{bench1} has a time complexity $\mathcal{O}(W^3 M^3)$ due to the expensive computation of the \ac{mc} via \eq{eq:coherence}.
Conversely, \ac{bench2} has complexity $\mathcal{O}(2^U U^2 M)$, where $U$ is the number of non-zero components in the spectrum at the previous iteration.
Finally, the proposed \ac{rl} algorithm has a time complexity of $\mathcal{O}(Y M)$, where $Y$ denotes the number of operations performed by the \ac{cnn} to select the samples in a processing window, which depends on the size of the learning architecture.

In \tab{tab:time}, we report the median time, in milliseconds, spent by each technique to select a single sample to be inserted into $\mathcal{M}$.
Random selection has the lowest complexity, spending about $0.03$~ms per sample.
The \ac{bench1} and \ac{bench2} have high computational costs due to the computation of the \ac{mc}: in the scenario with \mbox{$M=8$}, they require more than $15$~ms and $46$~ms per action, respectively.
Appreciably, \ac{ppo} strongly outperforms the benchmarks and requires less than $2$~ms per sample in all configurations. 
This confirms how adopting data-driven approaches for \ac{cs}-based \ac{md} reconstruction is beneficial in terms of both accuracy and computational efficiency.} 

\section{Concluding remarks}

In this letter, we proposed a deep \ac{rl}-based approach to select sensing packets transmission times in communication-centric \ac{isac} for \acf{md} reconstruction.
Our framework sequentially selects the transmission times based on the available channel samples from communication packets and the last estimation of the \ac{md} spectrum.
The framework is trained with the \ac{ppo} algorithm to directly maximize the reconstruction quality, rather than the widely used \ac{mc} metric from the literature.
When tested on a real dataset of channel measurements, our system achieves up to $40$\% higher \ac{md} reconstruction accuracy and several times lower computational complexity compared to existing methods. 

\begin{table}[t!]
\centering
\footnotesize
\begin{tabular}{ccccc}
\toprule
$(M, b, d)$  & \textbf{PPO} & \textbf{RMC-RS} \cite{song2024nonuniform} & \textbf{MC-SFS} \cite{song2024nonuniform}& \textbf{Random} \\
\midrule
(8, 4, 1)  & 1.94    & 46.34    & 10.27    & 0.03      \\
(8, 4, 4)  & 1.95    & 46.28    & 10.20    & 0.03      \\
(8, 8, 1)  & 1.93    & 46.27    & 10.17    & 0.02      \\
(8, 8, 4)  & 1.94    & 46.17    & 10.16    & 0.03      \\
\midrule
(16, 8, 2)  & 1.89    & 15.54    & 9.37     & 0.03      \\
(16, 8, 8)  & 1.90    & 15.53    & 9.45     & 0.03      \\
(16, 16, 2) & 1.90    & 15.49    & 9.41     & 0.02      \\
(16, 16, 8) & 1.92    & 15.50    & 9.38     & 0.02      \\
\bottomrule
\end{tabular}
\caption{Median time (in milliseconds) spent by each algorithm to identify a new candidate to be added to set $\mathcal{M}_s$.}
\label{tab:time}
\end{table}

\bibliography{references}

\begin{thebibliography}{10}

\bibitem{dong2022sensing}
F.~Dong, F.~Liu, Y.~Cui, W.~Wang, K.~Han, and Z.~Wang, ``{Sensing as a service
  in 6G perceptive networks: A unified framework for ISAC resource
  allocation},'' {\em IEEE Trans. Wirel. Commun.}, vol.~22, pp.~3522--3536, May
  2023.

\bibitem{pegoraro2022sparcs}
J.~Pegoraro, J.~O. Lacruz, M.~Rossi, and J.~Widmer, ``{SPARCS: A Sparse
  Recovery Approach for Integrated Communication and Human Sensing in mmWave
  Systems},'' in {\em 21st ACM/IEEE International Conference on Information
  Processing in Sensor Networks (IPSN)}, (Milan, Italy), 2022.

\bibitem{vandersmissen2018indoor}
B.~Vandersmissen, N.~Knudde, A.~Jalalvand, I.~Couckuyt, A.~Bourdoux,
  W.~De~Neve, and T.~Dhaene, ``{Indoor person identification using a low-power
  FMCW radar},'' {\em IEEE Trans. Geosci. Remote Sens.}, vol.~56, no.~7,
  pp.~3941--3952, 2018.

\bibitem{zhang2021enabling}
J.~A. Zhang, M.~L. Rahman, K.~Wu, X.~Huang, Y.~J. Guo, S.~Chen, and J.~Yuan,
  ``{Enabling joint communication and radar sensing in mobile networks—A
  survey},'' {\em IEEE Commun. Surv. Tutor.}, vol.~24, pp.~306--345, January
  2022.

\bibitem{eldar2012compressed}
Y.~C. Eldar and G.~Kutyniok, {\em {Compressed sensing: theory and
  applications}}.
\newblock {Cambridge University Press}, 2012.

\bibitem{xu2014compressed}
G.~Xu and Z.~Xu, ``{Compressed sensing matrices from Fourier matrices},'' {\em
  IEEE Trans. Inf. Theory}, vol.~61, pp.~469--478, January 2015.

\bibitem{song2024nonuniform}
Z.~Song, Y.~She, J.~Yang, J.~Peng, Y.~Gao, and R.~Tafazolli, ``{Nonuniform
  Sampling Pattern Design for Compressed Spectrum Sensing in Mobile Cognitive
  Radio Networks},'' {\em IEEE Trans. Mob. Comput.}, vol.~23, pp.~8680--8693,
  September 2024.

\bibitem{pegoraro2022disc}
J.~Pegoraro, J.~O. Lacruz, M.~Rossi, and J.~Widmer, ``{DISC: a dataset for
  integrated sensing and communication in mmWave systems},'' {\em IEEE Dataport
  https://dx.doi.org/10.21227/2gm7-9z72}, 2022.

\bibitem{pegoraro2023rapid}
J.~Pegoraro, J.~O. Lacruz, F.~Meneghello, E.~Bashirov, M.~Rossi, and J.~Widmer,
  ``{RAPID: Retrofitting IEEE 802.11ay Access Points for Indoor Human Detection
  and Sensing},'' {\em Trans. on Mob. Comp.}, vol.~23, no.~5, pp.~4501--4519,
  2024.

\bibitem{sutton2018reinforcement}
R.~S. Sutton and A.~G. Barto, {\em Reinforcement learning: An introduction}.
\newblock MIT press, 2018.

\bibitem{schulman2017proximal}
J.~Schulman, F.~Wolski, P.~Dhariwal, A.~Radford, and O.~Klimov, ``{Proximal
  Policy Optimization Algorithms},'' {\em https://arxiv.org/abs/1707.06347},
  2017.

\bibitem{kingma2017adammethodstochasticoptimization}
D.~P. Kingma and J.~Ba, ``Adam: A method for stochastic optimization,'' {\em
  https://arxiv.org/abs/1412.6980}, 2017.

\end{thebibliography}
\bibliographystyle{ieeetr}
\end{document}